\documentstyle[11pt]{article}
\begin{document}

\title{\textbf{Ghost Chaplygin scalar field model of dark energy}}

\author{F. Adabi$^{1}$, K. Karami$^{2}$\thanks{KKarami@uok.ac.ir} , M. Mousivand${^2}$\\$^{1}$\small{Department of Physics, Sanandaj Branch, Islamic Azad
University, Sanandaj, Iran}\\$^{2}$\small{Department of Physics,
University of Kurdistan, Pasdaran St., Sanandaj, Iran}\\}

\maketitle

\begin{abstract}
We investigate the correspondence between the ghost and Chaplygin
scalar field dark energy models in the framework of Einstein
gravity. We consider a spatially non-flat FRW universe containing
the interacting dark energy with dark matter. We reconstruct the
potential and the dynamics for the Chaplygin scalar field model
according to the evolutionary behavior of the ghost dark energy
which can describe the phantomic accelerated expansion of the
universe.
\end{abstract}

\clearpage
\section{Introduction}
Observational data of type Ia supernovae (SNeIa) collected by Riess
et al. \cite{riess} in the High-redshift Supernova Search Team and
by Perlmutter et al. \cite{perl} in the Supernova Cosmology Project
Team independently reported that the present observable universe is
undergoing an accelerated expansion phase. The exotic source for
this cosmic acceleration is generally dubbed ``dark energy'' (DE).
Despite many years of research (see e.g., the reviews
\cite{Padmanabhan,Copeland}) its origin has not been identified till
yet. DE is distinguished from ordinary matter (such as baryons and
radiation), in the sense that it has negative pressure. This
negative pressure leads to the accelerated expansion of the universe
by counteracting the gravitational force. The astrophysical
observations show that about 73\% of the present energy of the
universe is contained in DE. There are several DE models to explain
cosmic acceleration.

One of interesting DE candidates is the Chaplygin gas which was
proposed to explain the accelerated expansion of the universe
\cite{Kamench}. The Chaplygin gas behaves as a pressureless dark
matter (DM) at early times and like a cosmological constant at late
stage \cite{Kamench}. This interesting feature leads to the
Chaplygin gas model being proposed as a candidate for the unified
DM-DE (UDME) scenario \cite{Wu}.

More recently, a new DE model called ghost DE (GDE) has been
motivated from the Veneziano ghost of choromodynamics (QCD). The
advantages of the GDE with respect to other DE models include the
absence of the fine tuning and cosmic coincidence problems and the
fact that it can be completely explained within the standard model
and general relativity, without recourse to any new field, new
degree(s) of freedom, new symmetries or modifications of general
relativity \cite{CaiGhost}. Several aspects of this new paradigm, in
particular observational constraints on this model, have been
investigated in the literature \cite{sheikh}.

The other interesting issue in modern cosmology is reconstructing
the scalar field models of DE which has been investigated in the
literature \cite{Zhang1}. The scalar field models (such as
quintessence, phantom, quintom, tachyon, K-essence, dilaton and
Chaplygin) are often regarded as an effective description of an
underlying theory of DE \cite{JPWu}. The scalar field models can
alleviate the fine tuning and coincidence problems \cite{Ali}.
Scalar fields naturally arise in particle physics including
supersymmetric field theories and string/M theory. Therefore, scalar
field is expected to reveal the dynamical mechanism and the nature
of DE. However, although fundamental theories such as string/M
theory do provide a number of possible candidates for scalar fields,
they do not uniquely predict its potential $V(\phi)$ \cite{JPWu}.
Therefore it becomes meaningful to reconstruct $V(\phi)$ from some
DE models (such as holographic, agegraphic and ghost).

All mentioned in above motivate us to investigate the correspondence
between the GDE and the Chaplygin scalar field model of DE. To do so
in Section 2, we study the interacting GDE with DM in a spatially
non-flat Friedmann-Robertson-Walker (FRW) universe. In Section 3, we
investigate the Chaplygin scalar filed model of DE. In Section 4, we
suggest a correspondence between the interacting GDE and the
Chaplygin scalar field model of DE. We reconstruct the potential and
the dynamics for the ghost Chaplygin scalar field model. Section 5
is devoted to conclusions.

\section{Interacting GDE with DM}
The GDE density is given by \cite{CaiGhost}
\begin{equation}\label{rhoGDE}
\rho_{\Lambda}=\alpha H,
\end{equation}
where $\alpha$ is a constant with dimension $[{\rm energy}]^3$, and
roughly of order of $\Lambda_{\rm QCD}^3$ where $\Lambda_{\rm
QCD}\sim100$ MeV is QCD mass scale. With $H\sim10^{-33}$ eV,
$\Lambda^3_{\rm QCD}H$ gives the right order of observed DE density.
This numerical coincidence is impressive and also means that this
model gets rid of fine tuning problem \cite{CaiGhost}.

Here, we investigate the GDE model in the framework of Einstein
gravity. To do so, we consider a spatially non-flat FRW universe
containing the GDE and DM. The first Friedmann equation in the
standard FRW cosmology is
\begin{equation}\label{1}
    H^{2}+\frac{k}{a^{2}}=\frac{1}{3M_{\rm p}^{2}}(\rho_{\rm m}+\rho_{\Lambda}),
    \end{equation}
where $M_{\rm p}=(8\pi G)^{-1/2}$ is the reduced Planck mass and
$H=\dot{a}/a$ is the Hubble parameter. Here, $k=0,1,-1$ represent a
flat, closed and open FRW universe, respectively. Also $\rho_{\rm
m}$
 and $ \rho_{\Lambda} $ are the energy densities of DM and GDE,
 respectively.

Using the usual definitions for the dimensionless energy densities
 as
\begin{equation}\label{2}
\Omega_{\rm m}=\frac{\rho_{\rm m}}{\rho_{\rm cr}}= \frac{\rho_{\rm
m}}{3M_{\rm p}^{2}H^{2}},~~~\Omega_{\Lambda}=
\frac{\rho_{\Lambda}}{\rho_{\rm cr}}= \frac{\rho_{\Lambda}}{3M_{\rm
p}^{2}H^{2}},~~~ \Omega_{k}=\frac{k}{a^{2}H^{2}},
\end{equation}
the Friedmann equation (\ref{1}) can be rewritten as
\begin{equation}\label{3}
1+\Omega_{k}=\Omega_{\rm m}+\Omega_{\Lambda}.
\end{equation}
Substituting Eq. (\ref{rhoGDE}) into $\rho_{\Lambda}=3M_{\rm
p}^2H^2\Omega_{\Lambda}$ yields
 \begin{equation}\label{OmegaL}
   \Omega_{\Lambda}=\frac{\alpha}{3M_{\rm p}^{2}H}.
 \end{equation}
Using the above relation, the curvature energy density parameter can
be obtained as
 \begin{equation}\label{omegak}
   \Omega_{k}=\left(\frac{9M_{\rm p}^{4}k}{\alpha^{2}}\right)\frac{\Omega_{\Lambda}^{2}}{a^2}
   =\left(\frac{\Omega_{k_0}}{\Omega_{\Lambda_0}^2}\right)\frac{\Omega_{\Lambda}^{2}}{a^2},
 \end{equation}
where we take $a_0=1$ for the present value of the scale factor.

We further assume there is an interaction between GDE and DM. The
recent observational evidence provided by the galaxy cluster Abell
A586 supports the interaction between DE and DM \cite{Bertolami8}.
In the presence of interaction, $\rho_{\Lambda}$ and $\rho_{\rm m}$
do not conserve separately and the energy conservation equations for
GDE and pressureless DM are
\begin{equation}\label{10}
        \dot{\rho}_{\Lambda}+3H(1+\omega_{\Lambda})\rho_{\Lambda}=-Q,
    \end{equation}
    \begin{equation}\label{11}
        \dot{\rho}_{\rm m}+3H\rho_{\rm m}= Q,
    \end{equation}
where $\omega_{\Lambda}=p_{\Lambda}/\rho_{\Lambda}$ is the equation
of state (EoS) parameter of the interacting GDE and $Q$ stands for
the interaction term. Following \cite{Pavon}, we shall assume
$Q=3b^{2}H\rho_{\Lambda}$ with the coupling constant $b^2$. This
expression for the interaction term was first introduced in the
study of the suitable coupling between a quintessence scalar field
and a pressureless DM field \cite{Zimdahl,Amendola}.

\begin{figure}
\includegraphics{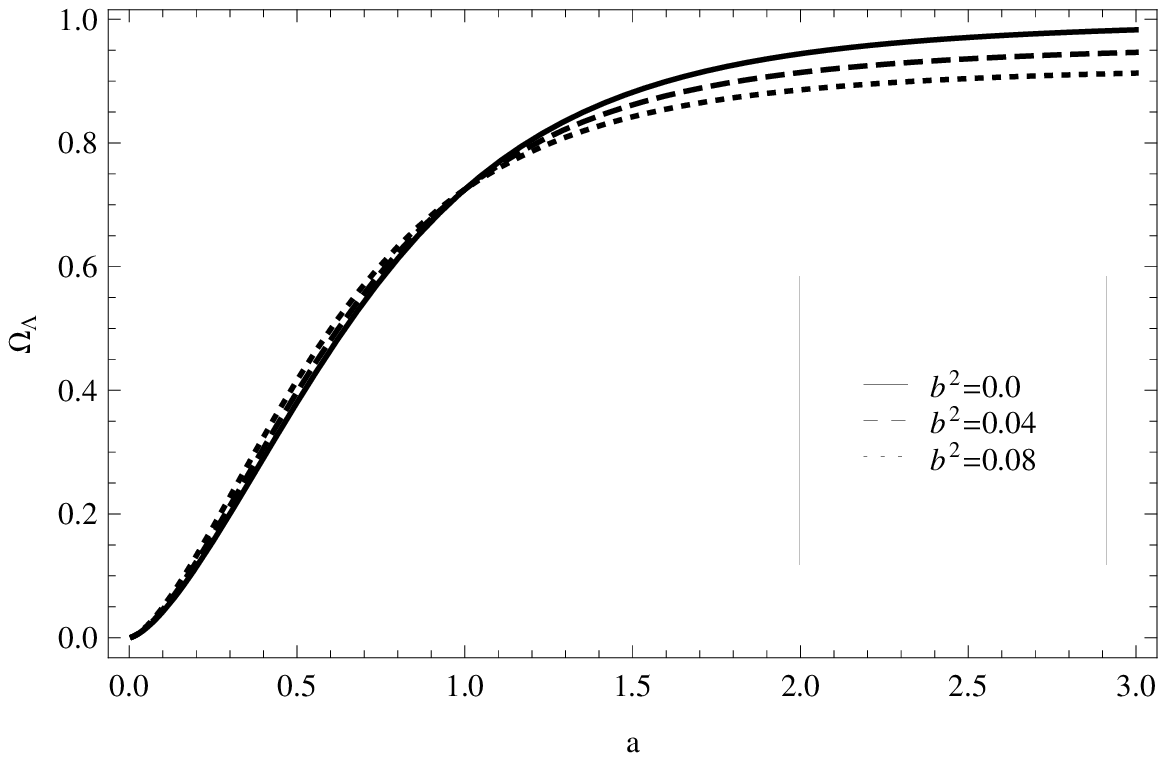}
      \vspace{4.7cm}
      \caption[]{The GDE density parameter, Eq. (\ref{diffOmega}), versus scale factor for different coupling constants
      $b^2=$ 0 (solid line), 0.04 (dashed line) and 0.08 (dotted line).
      Auxiliary parameters are $\Omega_{\Lambda_0}=0.725$ and $\Omega_{k_0}=0.0125$ \cite{Komatsu}.}
         \label{OmegaD-a}
   \end{figure}

Taking time derivative of Eq. (\ref{rhoGDE}) and using (\ref{1}),
(\ref{3}), (\ref{OmegaL}) and (\ref{11}) gives
\begin{equation}\label{7}
    \frac{\dot{\rho_{\Lambda}}}{\rho_{\Lambda}}
    =-\frac{3H}{2-\Omega_{\Lambda}}\left[1+\frac{\Omega_{k}}{3}-(1+b^2)\Omega_{\Lambda}\right].
\end{equation}
Taking time derivative of Eq. (\ref{OmegaL}) and using
(\ref{rhoGDE})
 and (\ref{7}) one can obtain the evolution
of the GDE density parameter as
\begin{equation}\label{diffOmega}
\frac{\Omega_{\Lambda}'}{\Omega_{\Lambda}}=\frac{3}{2-\Omega_{\Lambda}}
\left[1+\frac{\Omega_{k}}{3}-(1+b^2)\Omega_{\Lambda}\right],
\end{equation}
where $\Omega_{\Lambda}'=\frac{{\rm d} \Omega_{\Lambda}}{{\rm d}\ln
a}=\dot{\Omega}_{\Lambda}/H$. Using Eq. (\ref{omegak}) and taking
$\Omega_{k_0}=0.0125$ \cite{Komatsu} for the present time, one can
obtain $\Omega_{\Lambda}(a)$ by solving the differential equation
(\ref{diffOmega}) numerically with the initial condition
$\Omega_{\Lambda_0}=0.725$ \cite{Komatsu}. The numerical results
obtained for $\Omega_{\Lambda}(a)$ are displayed in Fig.
\ref{OmegaD-a} for different coupling constants $b^2$. Figure
\ref{OmegaD-a} shows that: i) for a given $b^2$, $\Omega_{\Lambda}$
increases when the scale factor increases. ii) At early and late
times, $\Omega_{\Lambda}$ increases and decreases with increasing
$b^2$, respectively.

With the help of Eqs. (\ref{10}) and (\ref{7}), the EoS parameter of
the interacting GDE can be obtained as
\begin{equation}\label{13}
    \omega_{\Lambda}=-\frac{1}{2-\Omega_{\Lambda}}
    \left(1-\frac{\Omega_{k}}{3}+2b^{2}\right).
\end{equation}
Equation (\ref{13}) shows that in the absence of interaction, i.e.
$b^2=0$, at late times where $ \Omega_{\Lambda}\rightarrow1 $ and
 $ \Omega_{k}\rightarrow0 $, we
have $\omega_{\Lambda}=-1 $ which behaves like $\Lambda$CDM.
Besides, taking $\Omega_{\Lambda_0}=0.725$ and $\Omega_{k_0}=0.0125$
for the present time then we obtain $\omega_{\Lambda_0}=-0.78$ which
acts like the quintessence DE ($\omega_{\Lambda}>-1$). But in the
presence of interaction, taking again $\Omega_{\Lambda_0}=0.725$ and
$\Omega_{k_0}=0.0125$ for the present time then from Eq. (\ref{13})
the EoS parameter can behave like phantom DE ($\omega_{\Lambda}<-1$)
provided $b^2>0.14$. This value for the coupling constant $b^2$ is
consistent with the observations \cite{Wang12}. Also the phantom
divide crossing is compatible with the recent observations
\cite{Komatsu}.

\begin{figure}
\includegraphics{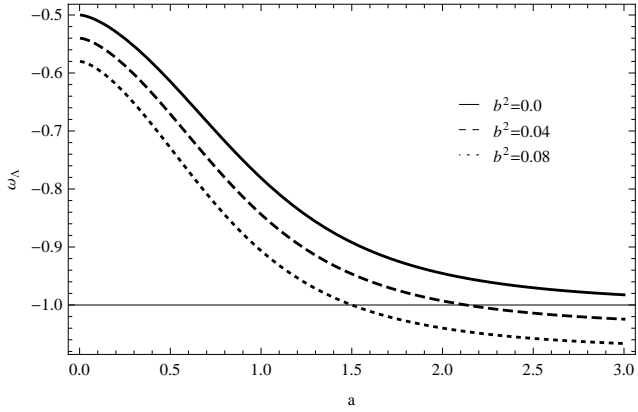}
      \vspace{4.7cm}
      \caption[]{The EoS parameter of GDE, Eq. (\ref{13}), versus scale factor.
      Legend and auxiliary parameters as in Fig. \ref{OmegaD-a}.}
         \label{wD-a}
   \end{figure}

The evolution of the EoS parameter, Eq. (\ref{13}), for different
$b^2$ is plotted in Fig. \ref{wD-a}. It shows that: i) for $b^2=0$,
$\omega_{\Lambda}$ varies from $\omega_{\Lambda}>-1$ to
$\omega_{\Lambda}=-1$, which is similar to the freezing quintessence
model \cite{Caldwell}. ii) For $b^2\neq 0$, $\omega_{\Lambda}$
varies from the quintessence phase ($\omega_{\Lambda}>-1$) to the
phantom regime ($\omega_{\Lambda}<-1$). iii) For a given scale
factor, $\omega_{\Lambda}$ decreases with increasing $b^2$.

\begin{figure}
\includegraphics{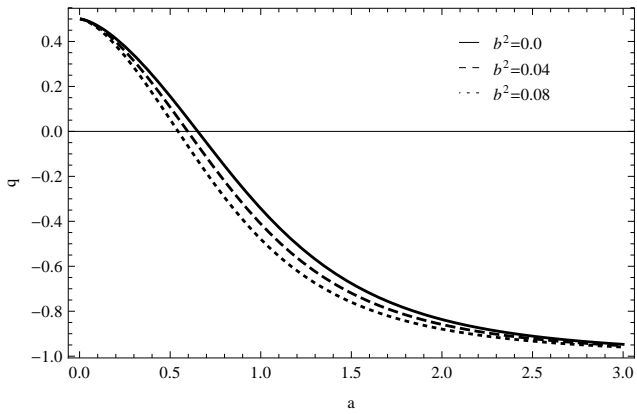}
      \vspace{4.7cm}
      \caption[]{The deceleration parameter, Eq. (\ref{q}), versus scale factor.
      Legend and auxiliary parameters as in Fig. \ref{OmegaD-a}.}
         \label{q-a}
   \end{figure}

For completeness, we give the deceleration parameter
\begin{equation}
q=-1-\frac{\dot{H}}{H^2},\label{q1}
\end{equation}
which combined with the dimensionless density parameter
$\Omega_{\Lambda}$ and the EoS parameter $\omega_{\Lambda}$ form a
set of useful parameters for the description of the astrophysical
observations. Using Eqs. (\ref{rhoGDE}), (\ref{7}) and (\ref{q1})
one can get
\begin{equation}
q=\frac{1+\Omega_k-(2+3b^2)\Omega_{\Lambda}}{2-\Omega_{\Lambda}}
  .\label{q}
\end{equation}
In Fig. \ref{q-a} we plot the evolutionary behavior of the
deceleration parameter, Eq. (\ref{q}), for different $b^2$. Figure
\ref{q-a} presents that: i) the universe transitions from a matter
dominated epoch at early times to the de Sitter phase, i.e. $q=-1$,
in the future, as expected. For $b^2=0.0$, $0.04$ and $0.08$ at
$a=0.65$, $0.60$ and $0.54$, respectively, we have a cosmic
deceleration $q>0$ to acceleration $q<0$ transition which is
compatible with the observations \cite{Ishida}. ii) For a given
scale factor, $q$ decreases with increasing $b^2$.

\section{Interacting Chaplygin scalar field with DM}
The EoS of the Chaplygin gas model of DE is as follows
\cite{Kamench}
\begin{equation}
p_{\rm Ch}=-\frac{A}{\rho_{\rm Ch}},\label{14}
\end{equation}
where $A$ is a positive constant. Inserting the above EoS into the
energy equation (\ref{10}) leads to a density evolving as
\begin{equation}\label{19}
   \rho_{\rm Ch}=\frac{1}{\sqrt{1+b^{2}}}\left(A +
   \frac{B}{a^{6(1+b^{2})}}\right)^{1/2},
\end{equation}
where $B$ is a positive integration constant. Note that the
Chaplygin gas model in the absence of interaction term, i.e.
$b^2=0$, offers a unified picture of DM and DE \cite{Wu}. Because it
smoothly interpolates between a non-relativistic matter phase
($\rho_{\rm Ch}\propto a^{-3}$) in the past and a negative-pressure
DE regime ($\rho_{\rm Ch}=-p_{\rm Ch}$) at late times.

Using Eqs. (\ref{14}) and (\ref{19}) the EoS parameter of the
interacting Chaplygin gas DE is obtained as
\begin{equation}
   \omega_{\rm Ch}=\frac{p_{\rm Ch}}{\rho_{\rm Ch}}=-1-\frac{Ab^{2}-Ba^{-6(1+b^{2})}}{A+Ba^{-6(1+b^{2})}},\label{wCG}
\end{equation}
which shows that for $Ab^{2}>Ba^{-6(1+b^{2})}$ we have $ \omega_{\rm
Ch} <-1 $ which corresponds to a universe dominated by phantom DE.
In other words, crossing the phantom divide line occurs when
\begin{equation}
a
> a_{\rm min}= \left(\frac{B}{Ab^{2}}\right)^{\frac{1}{6(1+b^{2})}}.
\end{equation}
Note that in the absence of interaction ($b^2=0$), Eq. (\ref{wCG})
gives $ \omega_{\rm Ch}
>-1 $ which corresponds to a universe dominated by quintessence DE.

Here, one can obtain a corresponding potential for the Chaplygin gas
DE by treating it as an ordinary scalar field $\phi$. Using Eqs.
(\ref{14}) and (\ref{19}) together with
 $ \rho_{\rm Ch}= \frac{1}{2}\dot{\phi}^{2}+V(\phi) $ and
$ p_{\rm Ch}= \frac{1}{2}\dot{\phi}^{2}-V(\phi) $, we find
\begin{equation}\label{21}
    \dot{\phi}^{2}=-\frac{Ab^{2}-Ba^{-6(1+b^{2})}}
    {(1+b^{2})^{\frac{1}{2}}\Big[A+Ba^{-6(1+b^{2})}\Big]^{\frac{1}{2}}},
\end{equation}
\begin{equation}\label{22}
   V(\phi)=\frac{A(2+b^2)+Ba^{-6(1+b^{2})}}
   {2(1+b^{2})^{\frac{1}{2}}\Big[A+Ba^{-6(1+b^{2})}\Big]^{\frac{1}{2}}}.
\end{equation}
Equation (\ref{21}) clears that for $Ab^{2}> Ba^{-6(1+b^{2})}$, the
Chaplygin gas DE behaves like a phantom scalar filed, i.e.
$\dot{\phi}^{2}<0$, whereas in the absence of interaction it acts
like a quintessence scalar field ($\dot{\phi^{2}}>0$).
\section{Correspondence between GDE and Chaplygin gas}
Here, our aim is to investigate whether a minimally coupled
Chaplygin scalar field can mimic the dynamics of the GDE model so
that this model can be related to some fundamental theory (such as
string/M theory), as it is for a scalar field. For this task, it is
then meaningful to reconstruct the $V(\phi)$ of Chaplygin scalar
field model possessing some significant features of the underlying
theory of DE, such as the GDE model. In order to do that, we
establish a correspondence between the GDE and Chaplygin gas scalar
field by identifying their respective energy densities and equations
of state and then reconstruct the potential and the dynamics of the
field.

Equating Eqs. (\ref{rhoGDE}) and (\ref{19}), i.e.
$\rho_{\Lambda}=\rho_{\rm Ch}$, gives
\begin{equation}\label{32}
   A=\alpha^{2}H^{2}(1+b^{2})-Ba^{-6(1+b^{2})}.
\end{equation}
Also equating Eqs. (\ref{13}) and (\ref{wCG}), i.e.
$\omega_{\Lambda}=\omega_{\rm Ch}$, and using (\ref{32}) we obtain
\begin{equation}\label{33}
   A=\frac{\alpha^{2}H^{2}}{2-\Omega_{\Lambda}}\left(1-\frac{\Omega_{k}}{3}+2b^{2}\right).
\end{equation}
Substituting Eq. (\ref{33}) into (\ref{32}) yields
\begin{equation}\label{34}
    B=\alpha^{2}H^{2}a^{6(1+b^{2})}\left[1+b^{2}-\frac{1}{2-\Omega_{\Lambda}}
    \left(1-\frac{\Omega_{k}}{3}+2b^{2}\right)\right].
\end{equation}
With the help of Eqs. (\ref{33}) and (\ref{34}) and using
(\ref{OmegaL}) one can rewrite (\ref{21}) and (\ref{22}) as
\begin{equation}\label{35}
    \dot{\phi^{2}}=\frac{\alpha^2}{3M_{\rm p}^2\Omega_{\Lambda}}\left[1-\frac{1}{2-\Omega_{\Lambda}}\left(1-\frac{\Omega_{k}}{3}+2b^{2}\right)\right],
\end{equation}
\begin{equation}\label{36}
   V(\phi)=\frac{\alpha^2}{6M_{\rm p}^2\Omega_{\Lambda}}\left[1+\frac{1}{2-\Omega_{\Lambda}}\left(1-\frac{\Omega_{k}}{3}+2b^{2}\right)\right].
\end{equation}
At the present time, if one takes $\Omega_{\Lambda}=0.73$ and
$\Omega_k=0.01$ then from Eq. (\ref{35}) one can obtain a phantom
scalar field ($\dot{\phi}^{2}<0$) provided $b^2>0.14$.

From definition $\dot{\phi}=\phi'H$ one can rewrite Eq. (\ref{35})
as
\begin{equation}\label{28}
    \phi'=\sqrt{3}M_{\rm p}\left\{\Omega_{\Lambda}\left[1-\frac{1}{2-\Omega_{\Lambda}}
    \left(1-\frac{\Omega_{k}}{3}+2b^2\right)\right]\right\}^{\frac{1}{2}},
\end{equation}
where $\phi'=\frac{{\rm d}\phi}{{\rm d}\ln a}$.

Using Eqs. (\ref{rhoGDE}) and (\ref{2}), the evolutionary form of
the Chaplygin gas scalar field can be obtained as
\begin{equation}\label{b}
   \phi(a)-\phi(a_0)=\sqrt{3}M_{\rm p}\int^{a}_{a_0}
\left[\frac{\Omega_{\Lambda}}{2-\Omega_{\Lambda}}\left(1-\Omega_{\Lambda}+
\frac{\Omega_{k}}{3}-2b^{2}\right)\right]^{1/2} ~\frac{{\rm
d}a}{a}.\label{phi1}
\end{equation}

\subsection{Numerical results}

\begin{figure}
\includegraphics{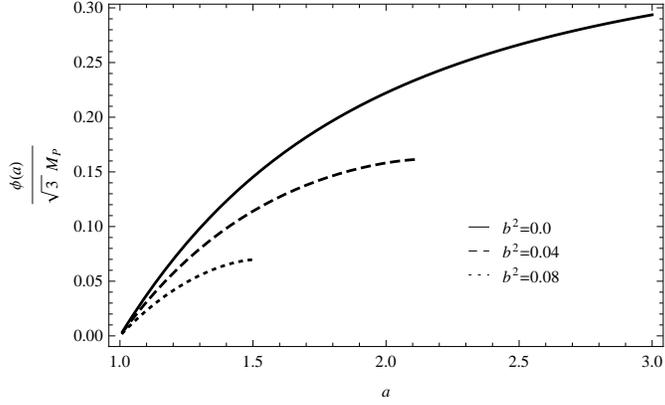}
      \vspace{4.7cm}
      \caption[]{The Chaplygin gas scalar field, Eq. (\ref{phi1}), versus scale factor.
      Legend and auxiliary parameters as in Fig. \ref{OmegaD-a}. Here $\phi(1)=0.0$.}
         \label{Phi-a}
   \end{figure}

The integral (\ref{phi1}) cannot be taken analytically. But with the
help of Eq. (\ref{omegak}) and numerical solution of the
differential equation (\ref{diffOmega}) one can take the integral
(\ref{b}), numerically. As we already mentioned, to solve Eq.
(\ref{diffOmega}) we need an initial condition for
$\Omega_{\Lambda}$ where we take its present value at $a_0=1$.
Besides, according to Eq. (\ref{omegak}) we need the present value
of $\Omega_{k_0}$. More recently, according to the 7-year WMAP data
\cite{Komatsu} the latest observational values of the aforementioned
parameters have been reported as $\Omega_{\Lambda_0}=0.725\pm 0.016$
and $\Omega_{k_0}=-0.0125_{-0.0067}^{+0.0064}$ at 68\% confidence
level. Note that in \cite{Komatsu} the curvature energy density
parameter is defined as $\Omega_{k}=-\frac{k}{a^{2}H^{2}}$ which
differs from our definition (\ref{2}) by a minus sign. Here, we set
the initial conditions to be the best fit values of
$\Omega_{\Lambda_0}=0.725$ and $\Omega_{k_0}=0.0125$. Also we take
$\phi(a_0)=0$ at present time ($a_0=1$). The evolution of the
Chaplygin gas scalar field (\ref{phi1}) is plotted in Fig.
\ref{Phi-a}. It shows that the Chaplygin scalar field increases when
the scale factor increases. Figure \ref{Phi-a} also clears that for
a given scale factor, the Chaplygin scalar field decreases with
increasing the coupling constant $b^2$. Note that in Fig.
\ref{Phi-a} for $b^2=0.04$ and $0.08$ at $a>2.13$ and $1.50$,
respectively, $\phi$ becomes pure imaginary, i.e. $\dot{\phi}^2<0$,
and does not show itself in Fig. \ref{Phi-a}. For $\dot{\phi}^2<0$
the Chaplygin scalar field behaves like phantom DE \cite{Caldwell2}.
Note that when the phantom scalar fields give rise to constant EoS
parameter $\omega_{\Lambda}$ smaller than $-1$, then the universe
reaches a Big Rip singularity in the future \cite{Narlikar}. In our
model according to Fig. \ref{wD-a}, for $b^2=0.0$, $0.04$ and $0.08$
when $a\rightarrow\infty$ we obtain $\omega_{\Lambda}=-1$, $-1.04$
and $-1.08$, respectively. This shows that in the presence of
interaction for $a\rightarrow \infty$ we have the constant EoS
parameter $\omega_{\Lambda}=p_{\Lambda}/\rho_{\Lambda}\simeq
-1-b^2<-1$ and fate of the universe goes toward a Big Rip.

\begin{figure}
\includegraphics{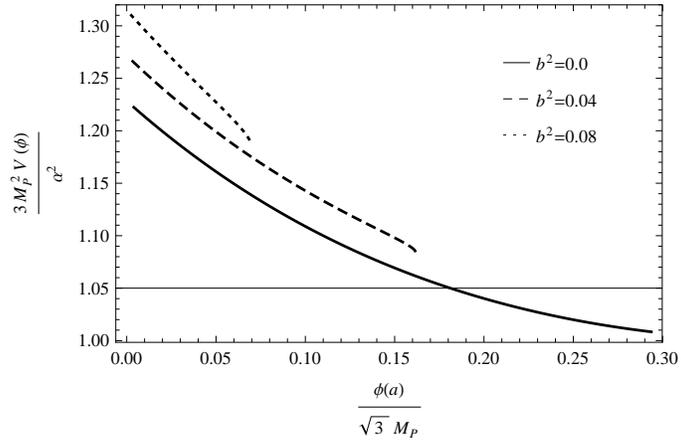}
      \vspace{4.7cm}
      \caption[]{The Chaplygin gas scalar potential, Eq. (\ref{36}), versus scalar field.
      Legend and auxiliary parameters as in Fig. \ref{OmegaD-a}.}
         \label{V-Phi}
   \end{figure}

The variation of the Chaplygin scalar potential (\ref{36}) versus
the scalar field $\phi$ is plotted in Fig. \ref{V-Phi}. It shows
that the Chaplygin scalar potential decreases when the scalar field
increases. Figure \ref{V-Phi} also presents that for a given scalar
field, the Chaplygin scalar potential increases with increasing the
coupling constant $b^2$.
\section{Conclusions}
Here we investigated the ghost Chaplygin scalar field model of DE in
the framework of FRW cosmology. We considered a spatially non-flat
FRW universe filled with the interacting GDE and pressureless DM. We
derived a differential equation governing the evolution of the GDE
density parameter. We also obtained the EoS parameter of the
interacting GDE and the deceleration parameter. Furthermore, we
established a correspondence between the GDE and the Chaplygin gas
scalar field model. We reconstructed the dynamics and the potential
of the Chaplygin gas scalar field according the evolutionary
behavior of the GDE model which can describe the phantomic
accelerated expansion of the universe at the present time. Our
numerical results show the following.

(i) The interacting GDE density parameter $\Omega_{\Lambda}$ for a
given coupling constant $b^2$, increases during history of the
universe. Also at early and late times, $\Omega_{\Lambda}$ increases
and decreases, respectively, with increasing $b^2$.

(ii) The EoS parameter $\omega_{\Lambda}$ of the GDE model can cross
the phantom divide line ($\omega_{\Lambda}<-1$) at the present
provided $b^2>0.14$ which is compatible with the observations. For
$b^2=0$, $\omega_{\Lambda}$ behaves like the freezing quintessence
DE. Also for $b^2\neq0$, $\omega_{\Lambda}$ varies from the
quintessence epoch ($\omega_{\Lambda}>-1$) to the phantom era.

(iii) The evolution of the deceleration parameter $q$ shows that the
universe transitions from an early matter dominant phase to the de
Sitter phase in the future which is in accordance with the
observations.

(iv) The ghost Chaplygin scalar filed $\phi$ for a given $b^2$,
increases with increasing the scale factor. Also for a given scale
factor, it decreases with increasing $b^2$. The ghost Chaplygin
potential $V(\phi)$ for a given $b^2$, decreases with increasing the
scalar filed. For a given scalar field, $V(\phi)$ increases with
increasing $b^2$.
\section*{Acknowledgements}
The authors thank the reviewers for a number of valuable
suggestions. The works of F. Adabi and K. Karami have been supported
financially by Department of Physics, Sanandaj Branch, Islamic Azad
University, Sanandaj, Iran.

\clearpage
\clearpage
\end{document}